\documentstyle[floats,prd,aps,epsf]{revtex} 
\draft 

\begin{document}

\twocolumn[\hsize\textwidth\columnwidth\hsize\csname
@twocolumnfalse\endcsname

\def\boxit#1{\vbox{\noindent\hrule\hbox{\vrule\hskip 3pt
\vbox{\vskip 3pt
\hbox{#1}\vskip 3pt}\hskip 3pt\vrule}\hrule}}

\overfullrule 0pt

\title{Head-On Collision of Neutron Stars As A 
Thought Experiment}

\author{Stuart L. Shapiro}

\address{Departments of Physics and Astronomy, \& NCSA, University of Illinois at 
Urbana-Champaign, Urbana, IL 61801}

\maketitle

\begin{abstract} 
The head-on collision of identical neutron stars from rest at infinity
requires a numerical simulation in full general relativity for a complete
solution. Undaunted, we provide a relativistic, analytic argument to suggest that
during the collision, sufficient thermal pressure is always generated 
to support the hot remnant in quasi-static stable equilibrium against collapse
prior to slow cooling via neutrino emission. Our 
conclusion is independent of the total mass of the progenitors and holds 
even if the remnant greatly exceeds the maximum mass of a cold neutron star.
\end{abstract}

\vskip2pc]

\section{Introduction}

Binary neutron stars are among the most promising sources
for gravitational wave laser interferometers now under construction,
like LIGO, VIRGO, GEO and TAMA. This fact has motivated an intense
theoretical effort to understand their dynamical evolution
and predict the gravitational waveforms emitted during their
inspiral and coalescence.

Here we focus on the final collision and coalescence of the
two stars.
The complex interplay between hydrodynamics and
gravitation, even in the idealized adiabatic case,
requires that this late epoch be tackled by numerical means.  
The numerical integrations must be carried out in full general
relativity even for a {\it qualitative}, let alone a quantitative, understanding of the
key phases of the final evolution. 
Strong-field phenomena characterizing the final stages of binary neutron 
star evolution, like gravitational collapse leading to black hole formation,  
are not addressed at all by Newtonian theory.

Consider the following three strong-field questions: Do binary neutron stars
collapse to black holes prior to contact? If not, does the remnant
undergo gravitational collapse immediately following the merger whenever its rest mass 
exceeds the maximum equilibrium mass of an isolated cold, rotating neutron star, or is
the collapse delayed ? 
What is the final fate of a collapsing remnant if its angular momentum exceeds the
Kerr limit, $J/M^2 \leq 1$ ? None of these questions can be 
answered via
a Newtonian simulation.
The answers have important consequences 
for the design of advanced detectors and theoretical templates for
measuring gravitational waves during the late epochs 
following collision, as well as other astrophysical consequences (e.g.,
gamma-ray burst models).

Here we address the second of the three questions posed above.
To simplify the analysis we treat the simplest merger scenario --    
the head-on collision of two identical, nonspinning, cold
neutron stars which fall together from rest at infinity.  The stars may have a combined
rest mass which greatly exceeds the hydrostatic equilibrium mass limit for an isolated,
nonrotating, spherical star governed by the same 
cold equation of state. Even this problem requires a numerical simulation in full
general relativity for a complete solution. However, by making a suitable set of 
simplifying approximations, 
we present an analytic argument to determine the fate of the remnant immediately
following merger. Our calculation may provide a
useful comparison for a $2 \mbox{D}$ or $3 \mbox{D}$ relativistic
code which can perform such a simulation. 

\section{Alternative Scenarios}

Following the head-on collision of the two neutron stars, 
two recoil shocks form at the point of contact and 
propagate back into each star along the collision axis. 
Matter passing through the shock is heated and decelerated, as its
kinetic energy is converted into thermal energy. 
The configuration is not in equilibrium
following the passage of the initial recoil shocks, but undergoes
oscillations, damped by acoustic (viscous)
dissipation and gravitational radiation reaction
forces.  Only a small fraction of the total
rest mass escapes from the system and an even smaller fraction of the total mass-energy
is carried off by gravitational waves. 

Two questions arise: Is the thermal pressure generated by shock 
and acoustic dissipation sufficient to support 
the merged remnant in hydrostatic equilibrium prior to cooling
by neutrino dissipation?  Does the answer depend on the total rest mass and its
relation to the maximum rest mass of an isolated, equilibrium mass supported by the
same initially cold equation of state? 

Consider two alternative scenarios. In the first we suppose that
the shock-generated thermal pressure is sufficiently large to prevent
the remnant from undergoing immediate collapse, allowing it instead to 
settle down to quasi-equilibrium in a few dynamical timescales ($\sim \mbox{msec}$) 
following
the initial oscillations. By then, the bulk of the kinetic 
energy will have been converted into thermal energy, the 
rest having been emitted as gravitational waves. The star will relax to a hot, 
quasi-static, spherical equilibrium state.
This equilibrium state will last many dynamical 
timescales because the thermal neutrinos,
which eventually carry off the thermal energy, leak out slowly, on a 
neutrino diffusion timescale ($\sim 10~\mbox{sec}$).
As the star cools, it slowly contracts. If its total rest mass is less than the
critical maximum value, it will evolve to a static, cold endpoint configuration
and cease all further contraction.  If its rest mass exceeds the critical value it must
ultimately undergo catastrophic collapse to a Schwarzschild black hole.
Since the quasi-static
configuration is spherical and  not too far from the Schwarzschild radius at the 
onset of this final `delayed' collapse, the implosion will remain nearly spherical and generate
little gravitational radiation. 

Now consider the alternative scenario, for which
the thermal pressure generated by shock heating is insufficient to
support the star in quasi-static equilibrium.  The star will then
collapse promptly to a black hole
on the relatively short dynamical timescale, and the bulk of the thermal energy and
neutrinos will go down the hole. 
Unable to achieve dynamical equilibrium, the 
compressed remnant  will 
never relax to a spherical shape so that its implosion will be
nonspherical. During the implosion, the mass may then emit a  
significant final pulse of gravitational radiation before forming a static Schwarzschild
black hole.

We show below that the first scenario is
energetically feasible, i.e, 
the merged remnant can reside in a hot, quasi-static 
stable equilibrium  phase lasting many dynamical 
timescales prior to cooling and contraction. Surprisingly, our 
conclusion is independent of the total rest 
mass of the object. Hence, even if the remnant mass greatly exceeds the maximum mass of a
cold neutron star, the quasi-equilibrium will be stable and the collapse
can be delayed considerably.

\section{Key Assumptions}

We assume that all of the original bulk kinetic energy at impact is 
dissipated into internal energy.  Our goal is
to assess whether the resulting 
pressure forces are capable of holding a nonrotating spherical remnant 
in stable equilibrium.
We treat the collision and merger adiabatically
and we assume that the combined rest mass and total mass-energy of the system is
strictly conserved. We neglect any net outflow of baryons from the
system, as well as the loss of energy via gravitational waves and neutrinos. 

We adopt an adiabatic equation of state in the form

\begin{equation}
\label{one}
P\ = K(s)\rho^\Gamma_0, 
\end{equation}
where $P$ is the pressure, $\rho_0$ is the rest-mass density, 
$\Gamma$ is the adiabatic index, assumed constant, 
and $K$ is a function of the specific entropy, $s$.
We assume that the approaching stars are cold with $s=0$
and take $K(0) = K_{\rm cold}$ to be a constant. We take the
cold equation of state which
supports the original neutron stars to be a  polytrope, with polytropic index $n$ given by
$\Gamma = 1 + 1/n$.
We assume that the value of $\Gamma$ defining the initial
polytropic equilibrium and the value associated with subsequent adiabatic changes
are identical, as is assumed in most hydrodynamic numerical simulations of
adiabatic gases. Adopting this assumption permits us to make 
a direct comparison with these simulations, as well as provide a
simplified analytic treatment of the collision. 
The correct equation of state of hot,
nuclear matter is still uncertain, and the actual variation of $\Gamma$ during
compression and shock heating is not known. We note that
$\Gamma$ remains strictly constant 
and equal to $5/3$ when ideal, degenerate, nonrelativistic neutrons
are heated to nonrelativistic temperatures by the shock. 
Following the passage of all
the shock and acoustic waves and the accompanying generation of entropy 
throughout
the matter, we approximate the function $K(s)$ appearing in Eq.~(\ref{one})
by a new constant, $K(s)=K_{\rm hot}$   
 ~\cite{ft2}.
We thus treat the merged remnant as a new, hot
spherical polytrope once it settles down.
Below we determine $K_{\rm hot}> K_{\rm cold}$  
uniquely in terms of $K_{\rm cold}$ for any $n$.

Some of the above assumptions can be calibrated
by comparing with existing Newtonian hydrodynamic
simulations  ~\cite{GS} ~\cite{RS} ~\cite{E}.  
The results of Rasio and Shapiro (hereafter RS; ~\cite{RS}),
who treated the adiabatic case $\Gamma=2$, are typical. They found 
that at most $5 \%$ of the total rest mass eventually escapes
and that the remnant settles 
down to a hot, nearly spherical equilibrium state.  
The loss of energy via gravitational waves is 
about $\Delta E/M = 2.4 \times 10^{-3} M_{1.4}R_{10}^{-7/2}$, where
$M_{1.4}$ is the mass of one star in units of $1.4 M_{\odot}$ and
$R_{10}$ is the radius in units of $10 ~\mbox{km}$. Hence, 
the assumptions of conserved rest-mass and total mass-energy 
during the initial dynamical epoch appear to be reasonable.

\section{Newtonian Analysis}

When two identical polytropes, each of (rest) mass $M_{\rm cold}$, radius
$R_{\rm cold}$ and gas constant $K_{\rm cold}$, collide head-on from 
rest at infinity, the total initial energy of the system (excluding rest-mass energy)
is given by

\begin{equation}
\label{three}
E_{\rm cold}\ = \ -\ {(3\Gamma -4)\over (5\Gamma -6)}\ \ {M^2_{\rm cold} 
\over R_{\rm cold}}
 \times\ 2.
\end{equation}
Here and throughout we set $G=c=1$.
The final remnant 
settles into a new polytropic configuration of the same index whose energy is given by
\begin{equation}
\label{four}
E_{\rm hot}\ =\ -\ {(3\Gamma -4) \over (5\Gamma -6)}\ \ {M^2_{\rm hot} 
\over R_{\rm hot}}.
\end{equation}
Conservation of rest-mass and total energy imply $M_{\rm hot}=2 M_{\rm cold}$ and
$E_{\rm hot}=E_{\rm cold}$ , which, together with Eqs.~(\ref{three}) and (\ref{four}),
require $R_{\rm hot}=2 R_{\rm cold}$.  The mass-radius relationship for
a polytrope,

\begin{equation}
\label{five}
M \ \propto \ R^{{3-n\over 1-n}}\ \ K^{{n\over n-1}}
\end{equation}
may then be used to determine $K_{\rm hot}$:

\begin{equation}
\label{six}
{K_{\rm hot}\over K_{\rm cold}}\ =\ \left({M_{\rm hot}\over M_{\rm cold}}\right)^{{n-1\over n}}
 \left({R_{\rm hot}\over R_{\rm cold}}\right)^{{3-n\over n}}\ \ =\ \ 4^{{1\over n}}.
\end{equation}

We test the validity of Eq.~(\ref{six}) by comparing with the
Newtonian simulations of RS, who monitor the time-evolution of the total
entropy for a $\Gamma=2$ head-on collision (see Fig 7 of their paper). 
Once the remnant settles down,
RS find that the entropy levels off at a value $S/(M_{\rm cold} k_{\rm B}/\mu) 
= 2\ln (K_{\rm hot}/K_{\rm cold}) \approx 3$.
Hence $K_{\rm hot}/K_{\rm cold} \approx 4.5$, which is in reasonable agreement
with the analytic prediction of Eq.~(\ref{six}),
$K_{\rm hot}/K_{\rm cold} \approx 4$. 

\section{General Relativistic Analysis}

Here we reconsider the same problem in the context of general relativity.
We again assume that the two polytropic stars collide head-on from 
rest at infinity, conserving both their total rest mass, $2M_0$ 
and total mass-energy, $2M$. 
At $t=0$ the two spherical stars are widely separated 
so that their mass-energies, like their
rest-masses, add linearly.  We show that
there exists a stable equilibrium solution for the remnant and determine the
value of $K_{\rm hot}$ corresponding to this solution.

Consider a sequence, parametrized by the
central density,  of spherical equilibrium stars 
constructed from the same one-parameter equation of state. 
The equilibrium models are solutions of the Oppenheimer-Volkoff (OV) equation of
hydrostatic equilibrium. Those solutions which are stable against radial
collapse reside on the ascending branch of the $M ~vs.~ \rho_c$ 
equilibrium curve, where $dM/d\rho_c > 0$. 
Each of our colliding neutron 
stars is situated on the stable branch of the OV spherical equilibrium curve for a polytrope
of constant $K=K_{\rm cold}$ at the
start of its infall.  We want to determine where the remnant resides on the corresponding
curve for a polytrope of constant $K = K_{\rm hot}$
once the stars merge and settle down (see Fig 1).

\begin{figure}
\epsfxsize=2.5in
\begin{center}
\leavevmode
\epsffile{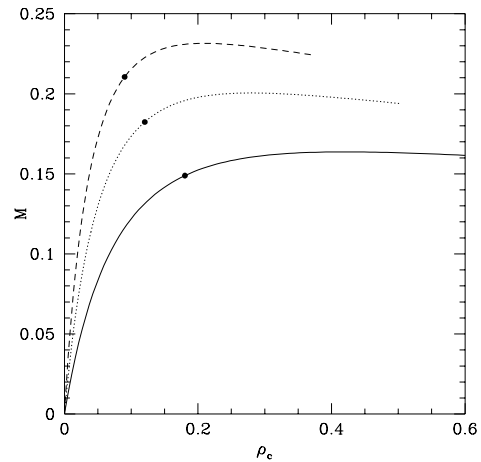}
\end{center}
\caption{Total mass-energy versus central density along 
the OV equilibrium curve for $n=1$ polytropes. The 
solid line shows models for $K=1$, the dotted line for 
$K=1.5$ and the dashed line for $K=2$. Solid points
locate representative stable models with the same universal
profile and binding energy.}
\end{figure}

The solutions exhibit a simple scaling behavior with respect
to $K$:

\begin{eqnarray}
\label{nine}
M(\rho_c)\ =\ K^{{n\over 2}} m^* (\rho^*_c)\ &,&\ \ M_0(\rho_c)\ =\ K^{{n\over
2}} m^*_0 (\rho^*_c)\nonumber \\
R(\rho_c)\ =\ K^{{n\over 2}} r^*(\rho^*_c)\ &,&\ \ \rho_c\ =\ K^{-n} \rho^{*}_c.
\end{eqnarray}
In Eq.~(\ref{nine}) $m^*, m_0^*$ and $r^*$ are 
universal, nondimensional
functions of the nondimensional central density $\rho_c^*$ and  depend
only on the index $n$. The universal equilibrium curves for an
$n=1$ polytrope are shown in Fig 2.

\begin{figure}
\epsfxsize=3in
\begin{center}
\leavevmode
\epsffile{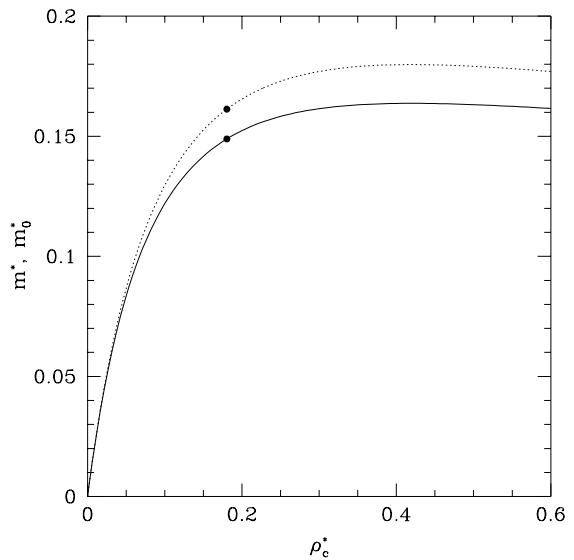}
\end{center}
\caption{Mass versus central density along the universal
OV equilibrium curve for an $n=1$ polytrope.
The solid line gives the total mass-energy, the dotted curve gives
the rest mass. Solid points locate the representative
stable model to which all the points in Fig 1 correspond.}
\end{figure}

\begin{figure}
\epsfxsize=2.5in
\begin{center}
\leavevmode
\epsffile{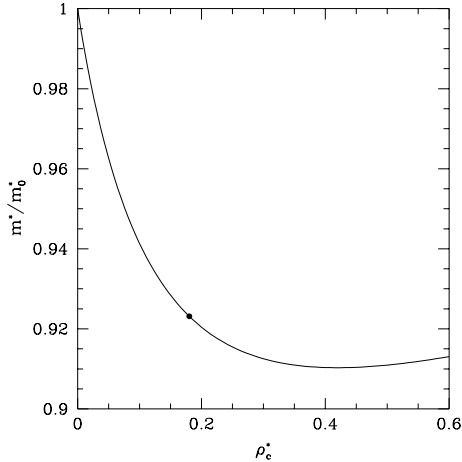}
\end{center}
\caption{Binding energy parameter versus central density along the
universal OV equilibrium curve for an $n=1$ polytrope. The solid point
locates the representative stable model to which all three points in
Fig 1 correspond.}
\end{figure}
The rest-mass and total mass-energy curves
have simultaneous turning points at $\rho_c^* =\rho_{\rm max}^*$, 
at which point 
both masses take on their
maximum values along the sequence.  The ratio
$m^*/m_0^*$ measures the specific binding energy of a star according to
$E_{\rm bind}/M_0=1-m^*/m_0^*$. This universal ratio 
monotonically decreases with $\rho_c^*$ along the stable branch
of the OV curve, $\rho_c^* \le \rho_{\rm max}^*$,
as shown in Fig 3. 
Conservation of rest mass and total mass-energy imply

\begin{equation}
\label{ten}
M_{0,{\rm hot}}\ =\ 2M_0\ ~\mbox{and}~\ M_{\rm hot}\ =\ 2M.
\end{equation}
Combining Eqs.~(\ref{nine}) and (\ref{ten}) gives
\begin{equation}
\label{eleven}
m^*_{0,{\rm hot}} / m^*_0  = m^*_{\rm hot} / m^* 
= 2 \left( K_{\rm hot} / K_{\rm cold} \right)^{-{n\over 2}},
\end{equation}
which implies
\begin{equation}
\label{twelve}
m^*_{\rm hot} / m^*_{0,{\rm hot}} = m^* / m^*_0.
\end{equation}
The nondimensional binding energy
ratio of the merged, hot remnant has the {\it same} value as in 
the cold progenitors. Because this ratio is a monotonic function of
$\rho_c^*$, the remnant can relax to the same stable, equilibrium point on 
the universal binding energy curve, hence

\begin{equation}
\label{thirteen}
m^*_{0,{\rm hot}}\ =\ m^*_0\ ~\rm{and}~\ m^*_{hot}\ =\ m^*.
\end{equation}
Thus, up to a scale factor $K$, the remnant profile is identical to that
of its progenitors.
Substituting Eq.~(\ref{thirteen}) back into ~(\ref{eleven}) implies

\begin{equation}
\label{fourteen}
{K_{\rm hot}\over K_{\rm cold}}\ =\ 4^{{1\over n}},
\end{equation}
whereby we recover the result 
obtained in the Newtonian analysis (cf. Eq.~(\ref{six}).

\section{Discussion and Caveats}

Head-on
collision and merger from rest at infinity need not lead to immediate collapse,
regardless of the mass of the incident stars. 
Sufficient thermal pressure will always be generated to create
a stable equilibrium configuration of higher entropy to 
which the fluid can can relax following merger: at fixed $\rho$

\begin{equation}
\label{fifteen}
P_{\rm therm} \approx P - P_{\rm cold}\ \approx\ P_{\rm cold}\ \left(4^{{1\over n}} -1\right),
\end{equation}
where $P_{\rm cold}$ is given by Eq.~(\ref{one}) with $K = K_{\rm cold}$ and
where

\begin{equation}
\label{sixteen}
P_{\rm cold} / \rho_{\rm cold} \sim\ M /  R.
\end{equation}
Setting the thermal pressure in the hot remnant
equal to the pressure of a Maxwell-Boltzmann
neutron gas
yields 

\begin{equation}
\label{seventeen}
k_BT \sim {{M m_n}\over R}\ {\rho_{\rm cold}\over\rho_{\rm hot}} 
\left(4^{{1\over n}} - 1\right)  \sim 140 \mbox{MeV} 
\left({M\over M_{\odot}}\right) \left({R\over 10\mbox{km}}\right)^{-1}.
\end{equation}
It is this thermal pressure that supports the remnant against 
collapse on a dynamical timescale ($\sim \mbox{msec}$). The dominant cooling mechanism 
is thermal neutrino radiation, but the timescale is comparatively long
($\sim 10~\mbox{sec}$) so the remnant evolves quasi-statically.

It is straightforward to modify
our argument to account for the escape of baryons, 
the loss of energy due to gravitational radiation, the
release of the progenitor stars from rest at finite separation rather than from infinity, 
etc.  In the later two circumstances we simply replace 
Eq.~(\ref{ten}) by

\begin{equation}
\label{eighteen}
M_{0,{\rm hot}}\ =\ 2M_0\ ~\rm{and}~\ M_{hot}\ =\ 2 ~ \epsilon ~ M,
\end{equation}
where $\epsilon \leq 1$ measures the decrease in the total mass-energy below the
initial value at infinite separation. Repeating the argument provides a 
constraint on $\epsilon$ for the remnant to relax to a
dynamically stable quasi-equilibrium state:

\begin{equation}
\label{nineteen}
{m^*_{\rm hot}\over m^*_{0,{\rm hot}}} = \epsilon  {m^* \over m^*_0} 
\geq  \left ({m^* \over m^*_0}\right )_{\rm min}
{\Longrightarrow  {\epsilon \geq  
{\left  ({m^* \over m^*_0}\right )_{\rm min} 
\over \left ({m^* \over m^*_0}\right )}}},
\end{equation}
where $(m^*/ m^*_0)_{\rm min}$ is the minimum value of the binding energy parameter
along the stable branch. This minimum occurs at the
maximum mass configuration and is $\approx 0.9$ for typical 
neutron stars (see Fig. 2).
The low expected radiation efficiency $f = 1 - \epsilon \ll 1$ 
~\cite{GS} ~\cite{RS} ~\cite{AE}
implies that all head-on collisions
from rest at infinity can lead to quasi-equilibrium except for those progenitors with masses
within $f$ of the maximum. For collisions starting from rest at finite separation
$d$, the total mass-energy is decreased by the initial binary interaction energy, so that

\begin{equation}
\label{twenty}
M_{\rm hot}\ \approx \ 2M \ - M^2 / d,~~~~~~~~(d \gg M).
\end{equation}
Equating Eqs. (\ref{eighteen}) and ~(\ref{twenty}) and ignoring radiation losses implies
$\epsilon \approx 1 - M/(2d)$, whereby the condition for stable equilibrium becomes

\begin{equation}
\label{twenty-one}
d \ \gtrsim \frac{M}{2} \left[ 1 - 
	\left({m^* \over m^*_0} \right)_{\rm min} 
	\left({m^* \over m^*_0} \right)^{-1} 
	\right]^{-1}.
\end{equation}
As $M \rightarrow  M_{\rm max}$
and $(m^*/m^*_0) \rightarrow (m^*/m^*_0)_{\rm min}$, we 
require that $d \rightarrow \infty$ for stable equilibrium
~\cite {ft4}. 

Simulations of head-on collisions of neutron stars in full
general relativity have been performed 
~\cite{AE} ~\cite{SW}, 
but no systematic studies have been published. The reported cases 
exhibit dynamical behavior qualitatively  similar to the Newtonian simulations 
regarding the recoil shocks, bounce and damped oscillations following the
collision, and there is no collapse  ~\cite{ft5}. 
This result is consistent with our theoretical expectations, but we 
await a more complete numerical investigation for confirmation.
There is the  possibility that the 
configuration  might greatly overshoot the final, allowed 
equilibrium state during the merger process and plunge immediately to
a black hole, but such a fate is not evident. Hopefully, this paper 
may stimulate further numerical experiments, 
including examples which employ a more realistic hot, nuclear equation of 
state. Off-axis collisions are also of interest; while shock heating will be 
less important in this case, the angular momentum acquired by the remnant may 
supplement the reduced thermal pressure to prevent sudden collapse
~\cite{BS}.
Resolving the fate of a head-on collision might serve
as a proving ground for a relativistic code designed to handle these more 
complicated 3D binary scenarios. 

\bigskip

\acknowledgements 

We thank A. Abrahams and T. Baumgarte for useful discussions.
This work has been supported in part
by  NSF Grant AST 96-18524 and NASA Grant NAG5-7152.


\begin{references}
\bibitem{ft2} The initial recoil shocks alone may 
generate nearly constant entropy when the equation of state is sufficiently stiff, 
in which case the progenitor stars are nearly homogeneous and the pre-shock gas parameters
nearly uniform.  Neutrino diffusion and convection may
further reduce the entropy gradient.
\bibitem{ft4} One must be especially careful to obey
constraint ~(\ref{twenty-one}) 
when setting up a numerical
simulation, where it is not possible to start
the stars off at infinite separation.
\bibitem{ft5} The gravitational waveform for the 
relativistic simulation reported in Refs ~\cite{AE} matches well with 
the Newtonian result reported in Ref ~\cite{RS}; see Fig 8 of
Ref ~\cite{RS}. The waveform probes the bulk motion
of the coalescing matter and therefore attests to the similarity of
the behavior.
\end{references}
\end{document}